# Ionic Behavior in Highly Concentrated Aqueous Solutions Nanoconfined between Discretely Charged Silicon Surfaces

Yinghua Qiu, Jian Ma and Yunfei Chen*

*Jiangsu Key Laboratory for Design and Manufacture of Micro-Nano Biomedical Instruments, School of Mechanical Engineering, Southeast University, Nanjing, 211189, China*

*Corresponding author:* yunfeichen@seu.edu.cn

*Tel: 86-138-158-888-16*

*Fax: 86-25-52090504*

**Abstract:**

Through molecular dynamics simulations considering thermal vibration of surface atoms, ionic behaviors in concentrated NaCl solutions confined between discretely charged silicon surfaces have been investigated. The electric double layer structure was found sensitive to the density and distribution of surface charges. Due to the surface charge discreteness, slight charge inversion appeared which depended on the surface charge density, bulk concentration and confinement. In the nanoconfined NaCl solutions differently concentrated from 0.2 M to 4.0 M, the locations of accumulation layers for $Na^+$ and $Cl^-$ ions kept stable, but their peak values increased. The higher the concentration was, the more obvious charge inversion appeared. In 4.0 M NaCl solution, $Na^+$ and $Cl^-$ ions show obvious alternating layered distributions which may be corresponding to the solidification found in experiments. By changing surface separation, the confinement had a large effect on ionic distributions. As both surfaces approached each other, many ions and water molecules were squeezed out of the confined space. Two adjacent layers in ion or water distribution profiles can be forced to closer to each other and merge together. From ionic hydration analysis, the coordination number of $Na^+$ ions in highly-confined space was much lower than that in the bulk.

## Introduction:

Concentrated solutions widely exist in biology, geology, physical chemistry and technology. At the solid-liquid interface, due to the electrostatic attraction by surface charges, counterions accumulate in the vicinity of the surface to form an electric double layer (EDL).[1] In order to better understand and control the EDL, an accurate description of ion distributions near charged surfaces is always pursued by scientists. When the bulk solution is dilute and the surface charge density is low, the Poisson-Boltzmann (PB) equation[1] can predict ion distributions in diffuse layers, which is a simple classical mean field theory and ignores the correlation among ions and surface charges. However, when the concentration is 0.1 M, the calculated Debye length[1] for a monovalent 1:1 solution is about 0.93 nm, which is close to the dimensions of hydrated ions. In the solution with a higher concentration above 0.1 M, the EDL is mainly described by the atomistic structure of the molecularly thin counterion layer. In this case, existing theories cannot describe the ion distribution accurately. Because the study of ionic distributions is of great importance in many domains, such as desalination or water purification, cosmetics, inks and paints, soil science and crystallization, DNA sequencing and electrical double layer capacitor (EDLC), ionic distributions deserve much attention.

Recent years have seen the experiments using advanced measurement instruments, including atomic force microscopy (AFM), surface force apparatus (SFA), and some other methods.[2-4] The force profiles in aqueous solution obtained by AFM or SFA are usually used to discover the chemical or physical properties at interfaces, such as ion distributions within the EDL and on surfaces,[5-9] surface potentials or surface charge densities,[10-14] as well as the ionic hydration performance.[15]

From the PB equation, a repulsive EDL force is usually produced when two similar charged objects approach each other, which composes the Derjaguin-Landau-Verwey-Overbeek (DLVO) force together with the van der Waals force.[16] With the increase of the bulk concentration, the surface charges are screened better by more concentrated counterions and the van der Waals force dominates the DLVO force. However, this attractive force between two similar charged surfaces[5] within ~5 nm was found suppressed with the concentration of monovalent aqueous solutions further increasing above ~1.0 M, such as NaCl, KCl, and CsCl.[12-13] The pair of solid surfaces could be silica-silica surfaces,[12] and silica-oxidized silicon surfaces,[13] which are negatively charged in contact with aqueous solutions. Dishon et al.[12] found that in NaCl, KCl and CsCl aqueous solutions pH 5.5, with the concentration increasing from 1 to 500 mM, the repulsion force between two silica surfaces obtained by AFM first was suppressed, then disappeared and finally reemerged. They thought the reemergence of the repulsion was because of surface charge inversion by excess adsorbed cations. Wang et al.[13] repeated this experiment and confirmed the results by measuring the forces between a silica sphere and an oxidized silicon wafer using the same solutions with concentrations ranging from 0.01 to 4.0 M. In a solution with the intermediate concentration, the attractive force appeared. With a further increase of the concentration, the attraction was suppressed, which they also assumed was due to the charge inversion.

At solid-liquid interfaces, because of the difference in physical and chemical properties between solid and liquid phases, as well as the unpredictable surface charge distribution, the water and ion distributions are very complicated. It has been confirmed that water molecules show a layered distribution within 1 nm away from the surface when the aqueous solution is dilute.[17-18] For ion distributions at an interface, the simplest case is the 2D distributions of counterions on charged surfaces, which determined the ionic structures in the normal direction. Through highly sensitive AFMs, the counterion distributions on charged surfaces have been found to largely depend on surface properties, especially the charge property.[9, 19] Also, the ionic adsorption to charged surfaces depends on the bulk concentration[20] and ionic hydration property.[6] When the distance between two charged surfaces is less than 3 nm, the hydration force plays an important role in the surface force, which can shed light on the water and ion structures, and lubrication[21-23] at interfaces. By frequency modulation AFM, Kilpatrick et al.[15] investigated the primary and structural hydration forces at mica-NaCl aqueous solution interface using an atomically sharp tip with the concentration ranging 10 to 500 mM. They found the decay length of the primary hydration force had a reduction with the concentration increase, but the force spacing and decay length of structural hydration remained constant. For the complicated ion distributions perpendicular to solid surfaces, layered ion structures have been found at solid-ionic liquid interfaces.[7] With concentrated aqueous solutions, Espinosa-Marzal et al.[5] found oscillatory force profiles resulting from ion-layering at concentrations below 100 mM, and Stern-layer solidification in ultra-highly concentrated $KNO_3$ solutions above 1 M confined between mica surfaces. They also guessed the hydration number in the highly confined electrolytes was much less than that in the bulk case. Their results were confirmed by Baimpos et al.[6] with CsCl and LiCl solutions of high concentrations through SFA and AFM. Above 0.1 M, both solutions exhibited oscillatory forces at the surface separation below 2.2 nm. And solidification of the inner EDL structure appeared in 3 M solutions.

In highly concentrated solutions, the correlation among ions and surface charges[5-6] cannot be neglected, which has large dependence on the surface charge density,[24] ionic valence[25] and size,[26] and bulk concentration.[12] Recent studies show that the correlation is of great importance and can influence the ion distributions at the solid-liquid interface.[27] Without considering the strong correlation, classical theories fail to predict the ion distributions at solid-liquid interfaces accurately, when the surface charge density is high or the solution is concentrated. Then, charge inversion appears.[27]

In the past two decades, multivalent counterions screening charged surfaces deserved lots of attractions. Many counterintuitive phenomena have been investigated, such as charge inversion[27] and like-charge attraction.[25] They not only challenge the conventional theories，but also suggest many new applications. For example, in biology, DNA molecules can have charge inversion in multivalent solutions and their effective surface charge becomes positive, which makes it convenient to conduct gene therapy at the negatively charged cell membrane.[28] Also, there have been many theoretical achievements for charge inversion and like-charge attraction, such as charge regulation theory,[1, 29] ion dispersion explanation[30] and strong correlated theory (SCT).[27] The most remarkable one is SCT, pioneered by Rouzina and Bloomfield,[31] and developed by Shklovskii[27] and Levin.[32] An essential ingredient is the formation of a 2D Wigner crystal of counterions on the charged surface, known as a strongly correlated liquid. The strongly correlated liquid provides a correlation-induced attraction force to pull many more counterions to attach to the surface, and leads to charge inversion. When several counterions aggregate near one surface charge, giant charge inversion can happen.[33-34] Using SCT, the net inverted charge and the critical concentration of cations in aqueous solution can be calculated for different surface geometries if the surface charge density and the bulk concentration are known.[27]

Due to the tiny scale, direct observations of ion and water distributions at interfaces are very difficult. Molecular dynamics (MD) simulations offer an effective method to discover the mysteries of the EDL. Some simulation results have confirmed the layered distribution of water molecules at solid-liquid interface.[24, 35] The orientation and concentration distribution profile of water molecules could depend on the surface crystal direction and charge property.[36-37] For the ion distribution in EDL, many simulations have been conducted using various models based on different solid materials, such as silicon[24, 36] and mica.[38] Due to the limitation of the simulation scale, concentrated solutions are usually used, such as 1 M NaCl or KCl. In order to obtain a higher energy density, highly concentrated solutions are widely used in EDLCs simulations.[39-40] However, few systematic researches have been conducted to explore the effect of concentration on the EDL, especially for solution concentrations above 1 M.

At solid-aqueous solutions interfaces, the solid surfaces are usually negatively charged due to ionization like mica and calcite, or dissociation of surface groups like silica.[1] In real systems the developed surface charges are elementary charges. The charges have a discrete character on solid surfaces, which is different from assumptions of the mean field theory, and can cause or enhance charge inversion.[41-43] Previously published MD simulations assumed uniformly distributed charges, which might not be the best representation of real systems.[24, 36] There are few simulations of the ion distributions at discreetly charged surfaces in highly concentrated solutions.[44] In addition, charge inversion phenomenon is seldom found for monovalent counterions in simulations. Using highly charged silicon surfaces to −0.285 C/m$^2$, Qiao et al.[24] found charge inversion and opposite electro-osmosis flow in 1M NaCl solution.

MD simulations were therefore conducted to detect the effect of surface charge density, solution concentration, and confinement on the EDL in highly concentrated monovalent solutions. The ion and water distributions in nanoconfined space between two silicon surfaces have been investigated using NaCl solutions. In our model, the thermal vibration of silicon atoms on solid walls was taken into consideration, which was always ignored in previous literatures.[24, 36] In efforts to emulate the real charged properties of the silicon surface, discrete surface charges were placed on the inner walls.[43] Our systems were not artificially electro-neutral, but were instead allowed to reach electrical neutrality through ion migration.[40, 43] We think our MD results in the vicinity of the charged surfaces can help to uncover the mystery of charge inversion and provide some advice to related domains.

## Method of Simulations

Figure 1 shows the schematic diagram of the simulation model, as we used before,[40, 43] which is assumed to be infinite in the *y* and *z* directions using periodic boundary conditions. A bulk and buffer region are used on both sides in *x* direction. Outside the bulk regions, a layer of sparse silicon atoms constitute a solid boundary to the aqueous solution. The middle part of the model is the nanoconfined space where the final results are obtained. Both the upper and bottom walls are composed of twelve-layer silicon atoms oriented in the (100) direction, of which eight-layer silicon atoms next to the solution are allowed to thermally vibrate near their equilibrium positions, while the outer four-layer silicon atoms are frozen without thermal vibration. In our MD model, the upper wall can move up and down to achieve different separations between the two surfaces. The bulk regions are used to provide enough ions and water molecules by filling with a predefined salt solution during the simulation. The buffer regions are used to judge whether the system achieves equilibrium. In the simulation, once the salt concentration, i.e. the ion number, in the buffer region does not change with time and keeps at the predefined value for enough long time, the system is believed to reach equilibrium. The following simulations are used to get the final results.

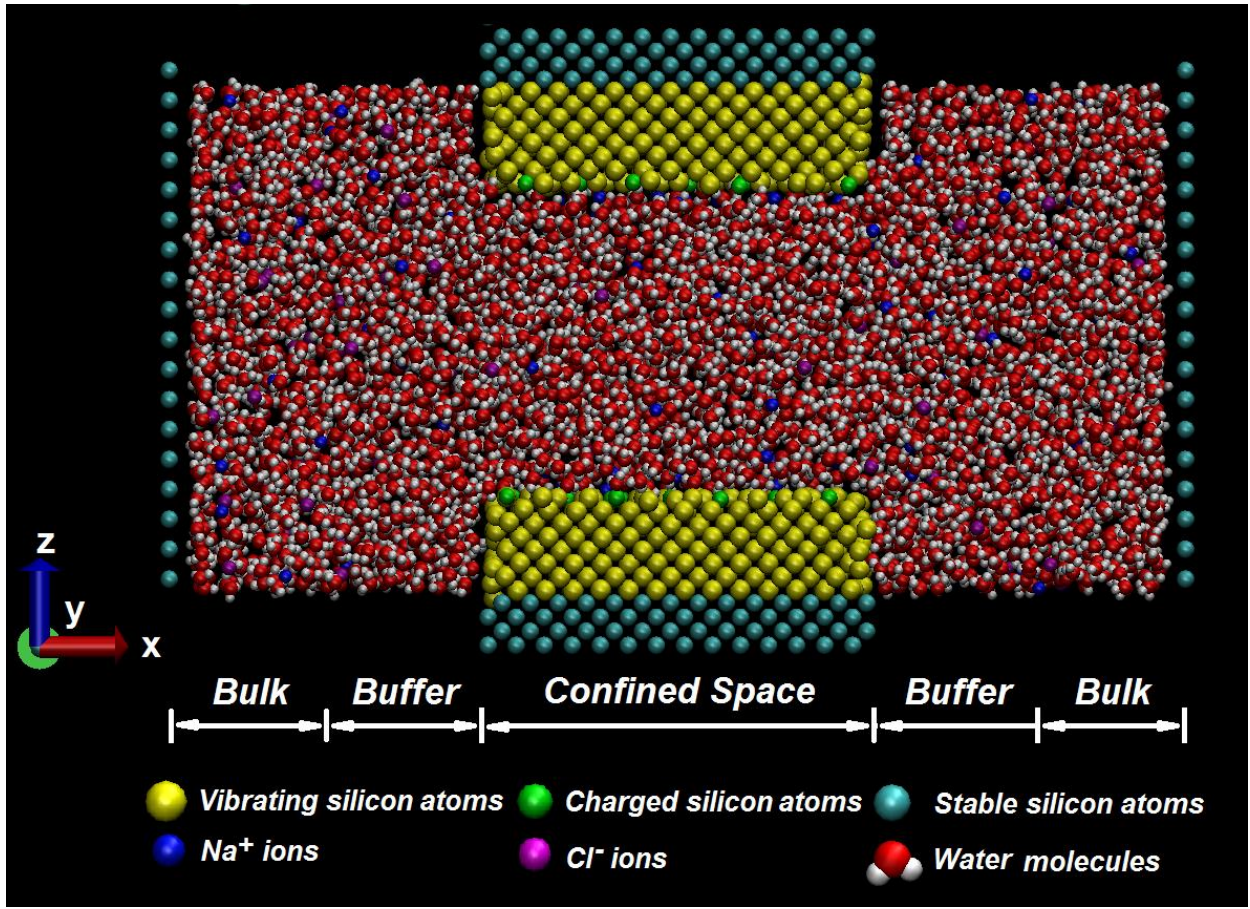

Fig. 1 Schematic diagram of the MD model. The dimensions of the system in x, y, and z directions are 10.2, 2.84, and 5.0 nm, respectively. The length of confined space is 3.8 nm.

At the beginning of the simulation, pure water was filled in the buffer and center regions and the NaCl solution with a predefined concentration was filled in the two bulk regions. Once the simulation started, ions would diffuse from the bulk regions to the confined space until the concentration gradients disappeared. The concentration gradients could be evaluated from the buffer regions. Before the system reached equilibrium, the solution in the two bulk regions was replaced periodically by the standard solution to provide enough ions. With two bulk regions, more ions could be supplied and the system can reach equilibrium much faster.[37]

In the simulation, discrete surface charges were used. One elementary charge was applied on each charged atom. The charged atoms were selected artificially to make them as uniformly distribute on the surface as possible. In the first part, different surface charge densities ranging from −0.03 C/m$^2$ to −0.3 C/m$^2$, as shown in Fig. 2, were used to detect their influence on ion and water distributions in 1 M NaCl solution. The distance between the two silicon surfaces was set at 3.0 nm based on the experimental results, because many interesting phenomena have been found when the surface separation is around 3 nm.[5-6] Due to the thermal vibration of the wall atoms, the actual average distance between the two charged surfaces is ~3.2 nm. In the second part, NaCl solutions with concentration from 0.2 to 4.0 M were filled in the systems with a 3.0 nm separation between upper and bottom walls. The surface charge density was set as −0.15 C/m$^2$. In the third part, the distance between the two walls was set as 1.7, 1.45, 1.25, 1.05, and 0.85 nm to detect the effect of confinement on ion distributions in 4.0 M NaCl solution. The surface charge density was −0.15 C/m$^2$. At the beginning of this series of simulations, the surface separation was 3.0 nm. Each confined case was achieved by compressing the nanospace i.e. moving down the upper wall rigidly. Then, the systems were allowed to reach equilibrium by replacing the bulk solutions periodically. The numbers of water molecules and ions used in the simulations are listed in Tables S1-S3.

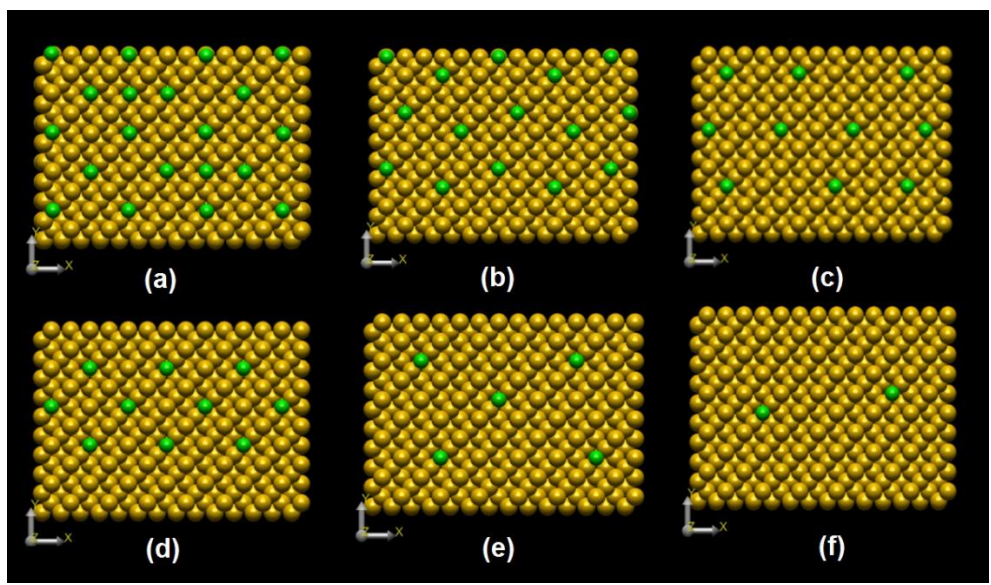

Fig. 2 Charge distributions on the surfaces. Yellow balls are uncharged silicon atoms, and green ones are charged. The charge numbers in (a)~(f) are 20,15,10 (Case 1),10(Case 2), 5, and 2, respectively.

The TIP4P model[45] was selected to simulate the water molecules. The SETTLE algorithm[46] was chosen to maintain the water geometry. LJ potentials [35] were used to describe the interactions between different atoms, except hydrogen-X pairs (X is an atom species in the solution) and silicon-silicon pairs (Table S4). The Stillinger-Weber (SW) potential[47] was used to describe the interaction among the silicon atoms. The electrostatic interactions among ions, water molecules and surfaces charges were modeled by Ewald summation algorithm.[48] The motion equations were integrated by the leap-frog algorithm with a time step of 2.0 fs. The solution system was maintained at 298 K by Berendesen thermostat[49] and the silicon walls were also maintained at 298 K using a damped method.[50] For each case, the first run lasting 8 ns was used to equilibrate the system. Another 6-ns-long run was followed to gather the statistical quantities. The concentrations of ions and water molecules were analyzed using the binning method.

## Results and discussions

1) **Different surface charge densities:**

Due to the electrostatic attraction, counterions accumulate in the vicinity of solid surfaces to screen the surface charges. Fig. 3(a) shows the counterion distributions perpendicular to the surfaces with different surface charge densities. When the surface is weakly charged to $-0.03$ C/m$^2$, a small accumulation layer of Na$^+$ ions appears at ~0.2 nm, which is close to its hydration radius.[51] This layer is identified as the outer Helmholtz layer in EDL.[52] As the surface charge density increases, many more Na$^+$ ions are attracted to the EDL. Until the surface is charged to $-0.15$ C/m$^2$, only outer Helmholtz layer exists in the EDL, whose location keeps stable though the electrostatic attraction enhances. However, when the surface charge density reaches $-0.225$ C/m$^2$, a new small peak corresponding to the inner Helmholtz layer (Fig. S1) appears nearer to the surface at 0.02 nm. In this layer the ions are non-hydrated or partly hydrated because few water molecules can approach so near to the surface (See below).[52] As the surface is charged further to $-0.3$ C/m$^2$, three peaks appear in the Na$^+$ ion distribution profile. A third peak located at 0.14 nm form between the inner and outer Helmholtz layers to screen the enhanced surface charges. In this case, the surface charge density i.e. $-0.3$ C/m$^2$ is very high but has been seen in some experiments at mica-aqueous solution interfaces.[53-54] From Fig. 3(a), it can be found that under the 5 different kinds of charged conditions, the Na$^+$ ions concentration drops to the bulk value sharply at 0.3 nm i.e. beyond the outer Helmholtz layer. Due to the thermal vibration of wall atoms, the solid surfaces are not atomically smooth. The strong interaction between the discreet surface charges and the counterions can

make some $Na^+$ ions penetrate into the rough silicon surfaces.[44]

For co-ions, they can hardly approach within 0.3 nm away from the surface, which is mainly because of the electrostatic repulsion from surface charges and the space occupation effects of counterions. Farther away from the surface, the concentration of $Cl^-$ ions increases sharply, and reaches its maximum value at ~0.62 nm. The peak position is close to the sum length of a hydration radius of $Cl^-$ ions and a hydration diameter of $Na^+$ ions.[51] The peak values increase with the surface charge density. At ~0.62 nm the $Cl^-$ ion concentrations exceed those of $Na^+$ ions[24, 27] which mean field PB theory cannot predict,[1] a phenomenon known as charge inversion appears. This is mainly attributed to that the strong correlation between surface charges and counterions make more $Na^+$ ions accumulate in the vicinity of the surface, which invert the effective surface charge from negative to positive. $Cl^-$ ions act as the counterions to screen the inverted surface charges i.e. the excess $Na^+$ ions (see below).

Within 1.0 nm of the surface, water molecules display an oscillatory distribution which is proved by some experiments[17-18, 55] and simulations.[24, 36] This is due to the attractive interactions between the solid surface and water molecules, and the geometric constraining effect of the surface on water molecules.[1] From Fig 3(c), the positions of corresponding peaks in each profile under a different charge condition remain the same. Farther away from 0.3 nm, the water distributions have nearly the same profile, which is not affected by the surface charge density. However, within 0.3 nm from the surfaces, the influence of surface charges cannot be ignored. With surface charge density increasing from −0.03 to −0.3 $C/m^2$, the adsorbed layer, which is next to the surface, locates nearer to the wall and accommodates more water molecules. A new water peak appears between the adsorbed layer and the hydration layer[18] when the surface is charged to −0.225$C/m^2$. It is caused by the polarity of water molecules and the enhanced attraction provided by the surface charges which produce an enlarged space between the adsorbed layer and the hydration layer to accommodate a new water layer. Our results about water molecule distributions show the same trend as the experimental results,[18] which are different from the earlier simulation data[24, 39] without thermal vibration of wall atoms.

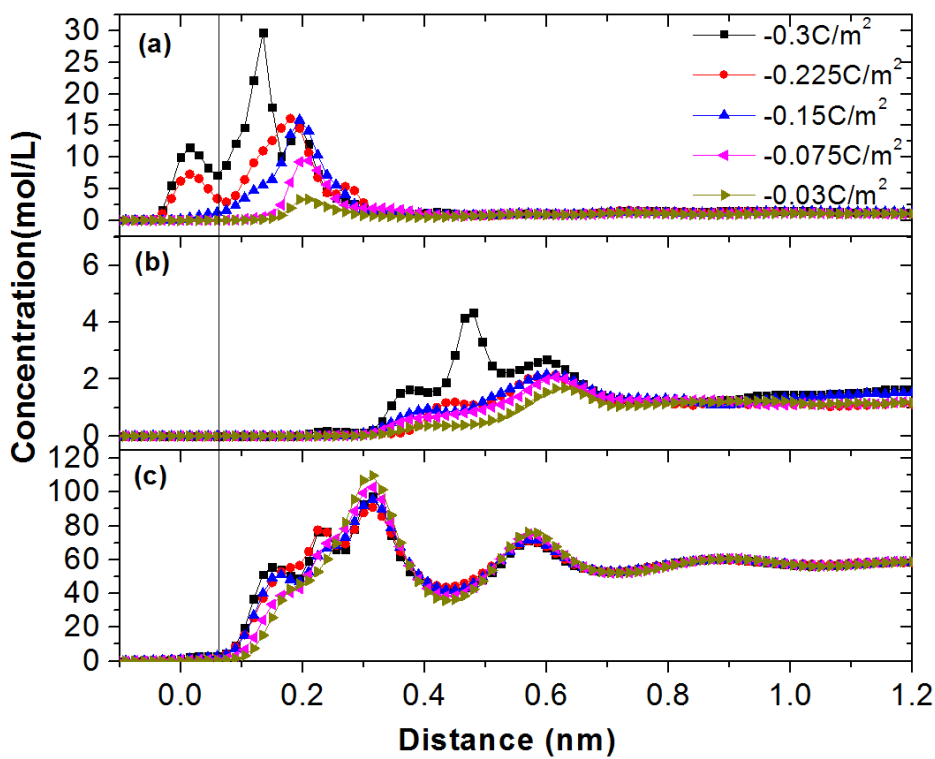

Fig 3. Ion and water concentration profiles perpendicular to the surface with charge density from −0.03 C/m$^2$ to −0.3 C/m$^2$ in 1 M NaCl solution at 3.0 nm separation between the surfaces. (a). Na$^+$ ions (b). Cl$^-$ ions (c). Water molecules. The corresponding charge distributions are shown in Fig 2(a), (b), (c), (e) and (f).

In the vicinity of charged surfaces, the appearance of the inner Helmholtz layer is caused by the enhanced surface charge density as shown in Fig. 3(a). In order to detect the effect of surface charge distribution on counterion distributions, a pair of simulations were conducted with equal surface charge density but different charge distributions, case 1 and case 2, shown in Fig.2 (c) and (d). In case 2, the surfaces have higher local charge density than those in case 1. The obtained Na$^+$ ions distributions are plotted in Fig. 4. It can be found that under the two different charged conditions, the main accumulation layers of Na$^+$ ions share the same position, 0.19 nm from the surface, due to the equal surface charge density. While, in case 2, a smaller peak forms next to the surface at 0.02 nm, which corresponds to the inner Helmholtz layer in the EDL theory and ion distributions in strongly charged cases. In our simulation system, no chemical interactions were considered. The appearance of the inner Helmholtz layer is caused by the electrostatic interactions between

ions and surface charges.[56]

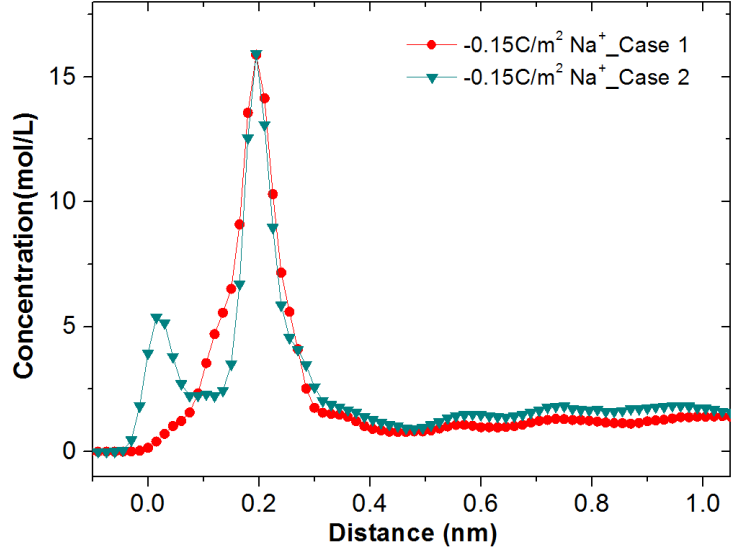

Fig. 4 Na$^+$ ions concentration profiles in two nanoconfined spaces with equal surface charge density but different charge distributions.

The integrated charge (IC) distribution perpendicular to the surface can be calculated through the following equation:[43]

$$\sigma_{IC}(i) = N_s + \sum_{t=1}^{i}\left[N_{Na^+}(t) + N_{Cl^-}(t)\right]$$

where $\sigma_{IC}(i)$ is the integrated charge accumulating to position $i$ from the surface, $N_s$ is the number of surface charges, $N_{Na^+}(t)$ and $N_{Cl^-}(t)$ are the local numbers of Na$^+$ and Cl$^-$ ions at position $t$.

The IC distributions near different charged surfaces are shown in Fig. 5(a), and the location of the surface charges is set as zero on the horizontal axis. Due to the formation of EDLs, the surface charges are screened by the counterions completely within ~0.25 nm from the surfaces which is close to the Debye length in 1 M NaCl.[1] Due to the surface charge discreteness, the strong correlation between surface charges and counterions attracts more Na$^+$ ions to the negatively charged surface which causes charge inversion.[27, 43] From the MD details (Fig. S2), two Na$^+$ ions were found nearby one surface charge, like several positive ions staying around one negative ion.[33-34] Though charge inversion happens in 1 M NaCl solution, it is much weaker than that in 0.5 M CaCl$_2$ solution (Fig. S3). This is because the correlation between surface charges and Ca$^{2+}$ ions are much stronger, which is consistent with the strong correlated theory.[27] Farther away from 0.25 nm, the effective surface charges change from negative to positive. In this case, Cl$^-$ ions act as the counterions to screen the overcrowded Na$^+$ ions, which causes the concentrations of Cl$^-$ ions exceed those of Na$^+$ ions in Fig. 3(b). In Fig. 5(b), the maximums of inverted charge at surfaces with different charge conditions are shown. With surface charge density increasing, the inverted charge amount increases, which is caused by the enhanced correlation between counterions and surface charges (Fig. S4).

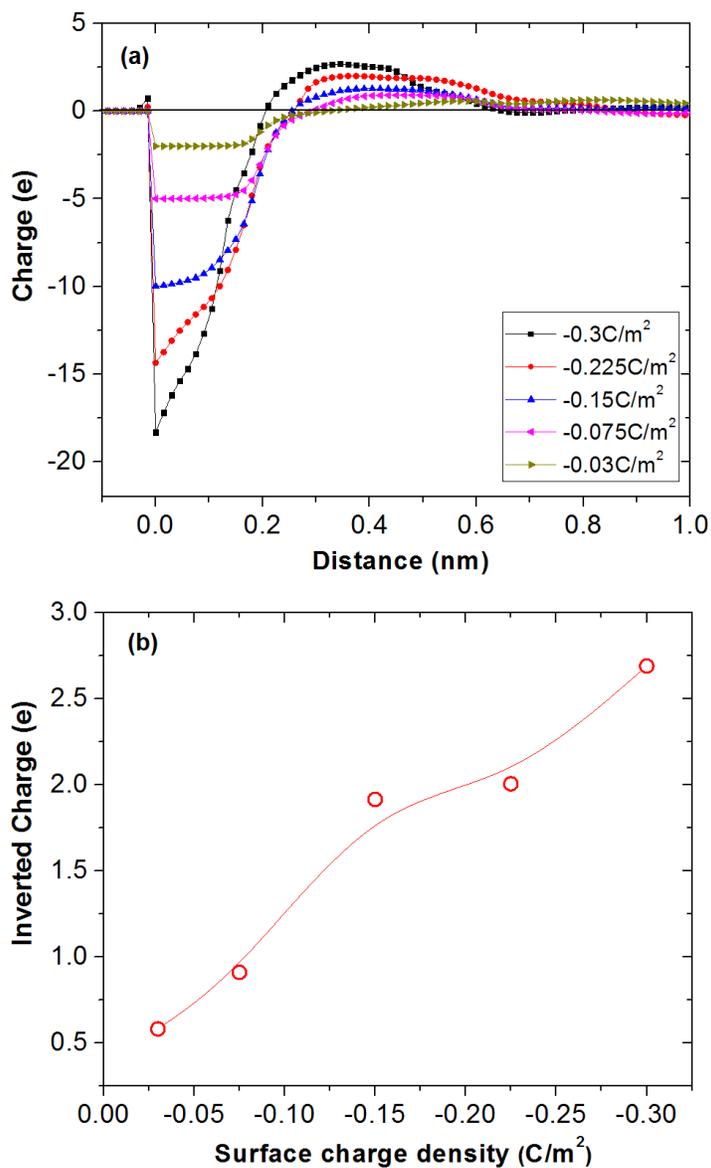

Fig. 5 (a) Integrated charge distributions as a function of the distance from the surfaces. The surface charges were located at 0 on the horizontal axis. (b) The maximums of inverted charge at each interface with different surface charge density from −0.03 C/m$^2$ to −0.3 C/m$^2$.

2) **Different solution concentrations:**

In order to investigate the influence of concentration on ion distributions in the confined spaces with highly concentrated monovalent aqueous solutions, a series of simulations were conducted using 0.2 to 4.0 M NaCl solutions. The separation between the surfaces was set as 3.0 nm and the surfaces were equally charged as −0.15 C/m$^2$ for each case. The surface charge distributions are shown in Fig. 2 (c).

The obtained ion distributions are shown in Fig. 6 (a). We can find that the peak positions of Na$^+$ and Cl$^-$ ions are not influenced by the concentration of solution, which locate at 0.19 nm and ~0.62 nm away from the surfaces. This is mainly due to the same charged solid surfaces in each case. However, with the concentration increase, both peak values of cations and anions are enhanced especially for anions, (Fig. S5) and more obvious charge inversion happens (see below): as a result, the effective surface potentials change from negative to positive. Cl$^-$ ions act as counterions to screen the charged surfaces. The increased peak values of Cl$^-$ ions can result in larger repulsive EDL force between two surfaces which should be responsible for the decreased attractive force and the reemergence of the repulsive force observed in the experiment with higher concentrations (Fig. S6).[12-13]

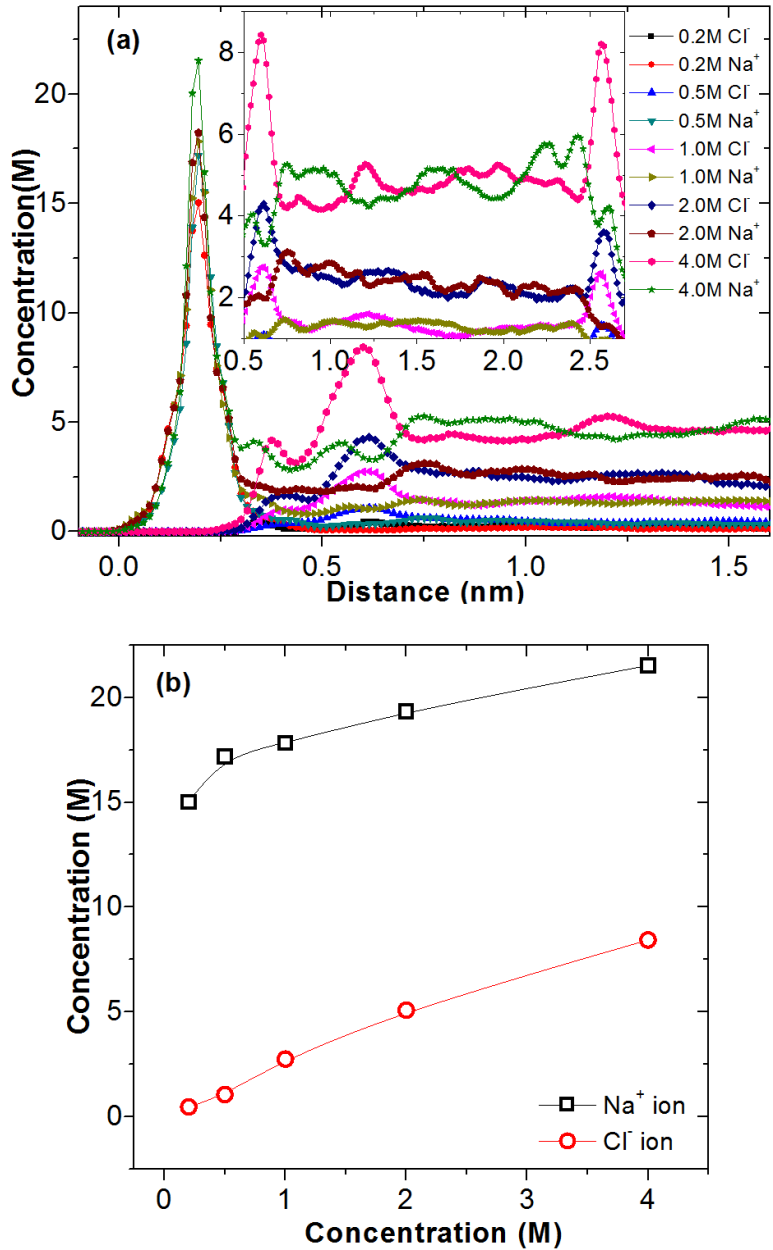

Fig. 6 (a) Ion concentration distributions in the normal direction of the same charged surfaces but immersed in differently concentrated NaCl solutions. The inset is the ion concentration profiles in the center of the nanospace. The surface charge density is −0.15C/m$^2$. (b) The peak values of Na$^+$ and Cl$^−$ ions which located at 0.19 nm and 0.62 nm away from the surface in NaCl solutions with different concentrations.

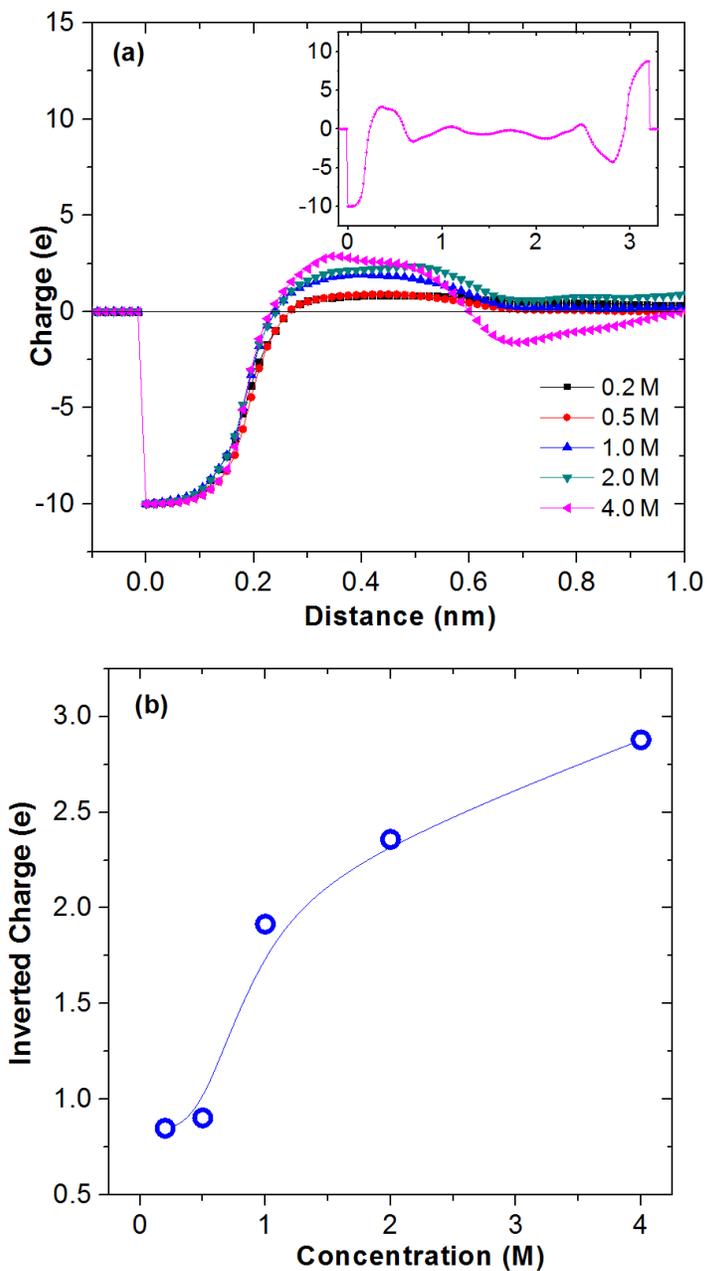

Fig. 7 (a) Integrated charge distributions in NaCl solutions with different concentrations. The inset is the whole distribution of the integrated charge in 4 M NaCl solution. (b)The inverted charge maximums at interfaces with differently concentrated NaCl solutions.

Figure 7(a) show the integrated charge distributions perpendicular to the surfaces immersed in NaCl solutions with different concentrations. The surface charges were set at 0 on the horizontal axis. It can be found that the surface charges are totally screened at around 0.25 nm from the surfaces, which is independent of the solution concentration. This is different from the Debye length (Fig. S8) and cannot be predicted by classical theory. Farther away, the accumulation of more counterions makes the solid surface show a inverted potential, which returns neutral at ~0.7 nm. The maximum of charge inversion in each case is shown in Fig. 7(b). As the bulk concentration increases, the correlation between surface charges and counterions enhances,[27] which reduces much more apparent charge inversion.

The ion distributions in the center of the nanospace are shown in the inset of Fig. 6(a). In NaCl solutions with concentration higher than 1 M, the cations and anions show alternating layered distribution between the solid surfaces, which is an ordered distribution style and similar to solidification in some extent.[5] The period is about 0.8 nm which is close to the sum of the hydration diameters of $Na^+$ and $Cl^-$ ions.[51] This phenomenon is found to be more obvious under higher surface charge density (Fig. S7) and higher confinement (see below).[40] Due to the obvious ionic layered distributions, the integrated charge profile in 4.0 M NaCl shows an oscillatory shape, as shown in the inset of Fig. 7(a). The layered distribution is mainly caused by the charge inversion and steric effect of hydration ions in confined space. In the simulations, a layer of $Na^+$ ions form to screen the negative surface charges. Due to the appearance of charge inversion, beyond the $Na^+$ ion layer, the effective surface potential is positive. A layer of $Cl^-$ ions appears to screen the over-crowded $Na^+$ ions. Because of the incompressibility of the hydrated ions, they have ordered distributions when large amount of them are highly confined between two charged surfaces.[1]

3) **Different confinements:**

A series of simulations with separation from 1.70 nm to 0.85 nm between two charged surfaces in 4.0 M NaCl solution were conducted to detect the effect of confinement on ion distributions in highly concentrated solutions. The counterion and co-ion distributions perpendicular to the surfaces are plotted in Fig. 8(a). We find that the counterions and co-ions show much more obvious layered distributions in more highly confined spaces (Fig. S9), which are similar to the ion distributions in ionic liquids[57] and our earlier results.[40] From Fig. 8(a), $Na^+$ ions mainly accumulate at their hydrated radius position, which is independent of the separation. As the upper wall approaches the bottom one, $Na^+$ ions show a double-layer distribution near the surface, a new layer emerges next to the main accumulation layer which is composed of the ions coming from the main accumulation layer (see below). When the separation is 0.85 nm, only one counterion layer is left because of the strongly restricted effect. In the center of all cases, $Na^+$ ions are found to distribute smoothly.

From Fig. 8(a), $Cl^-$ ions mainly accumulate in the center of the nanospaces. Due to the electrostatic repulsion of the surface charges, they appear at ~0.26 nm away from the surface where the surface potential has been almost screened by $Na^+$ ions. Beyond there, its concentration increases fast and exceeds that of $Na^+$ ions, i.e. charge inversion happens. With the increase of confinement, many $Cl^-$ ions are squeezed out and the location that $Cl^-$ ions can reach is compressed. In the 1.7 nm case, there are two obvious $Cl^-$ ions layers between the surfaces. However, as the separation decreases, both accumulation layers locate much nearer to each other and merge into one layer when the distance shrinks to 1.25 nm. With a further increase in confinement, fewer $Cl^-$ ions can stay between the charged surfaces.

Figure 8(b) shows the dependence of inverted charge on the surface separation obtained from the integrated charge distributions (Fig. S10). As the separation decreases, the inverted charge is diminished due to the enhanced confinement which squeezes more counterions from EDL and causes less counterions approach the charged surfaces. We think as the two charged surfaces approach each other further, there is a critical separation under which $Na^+$ ions will screen the surface charges exactly and there is no charge inversion. Higher confinement case (0.65 nm and 0.4 nm) have also been simulated, however we found that the surface charges cannot be totally screened by the counterions in the nanospaces. This may be due to the constant surface charge density we used in the simulations which decreases with the surface separation shrinking[58-59]

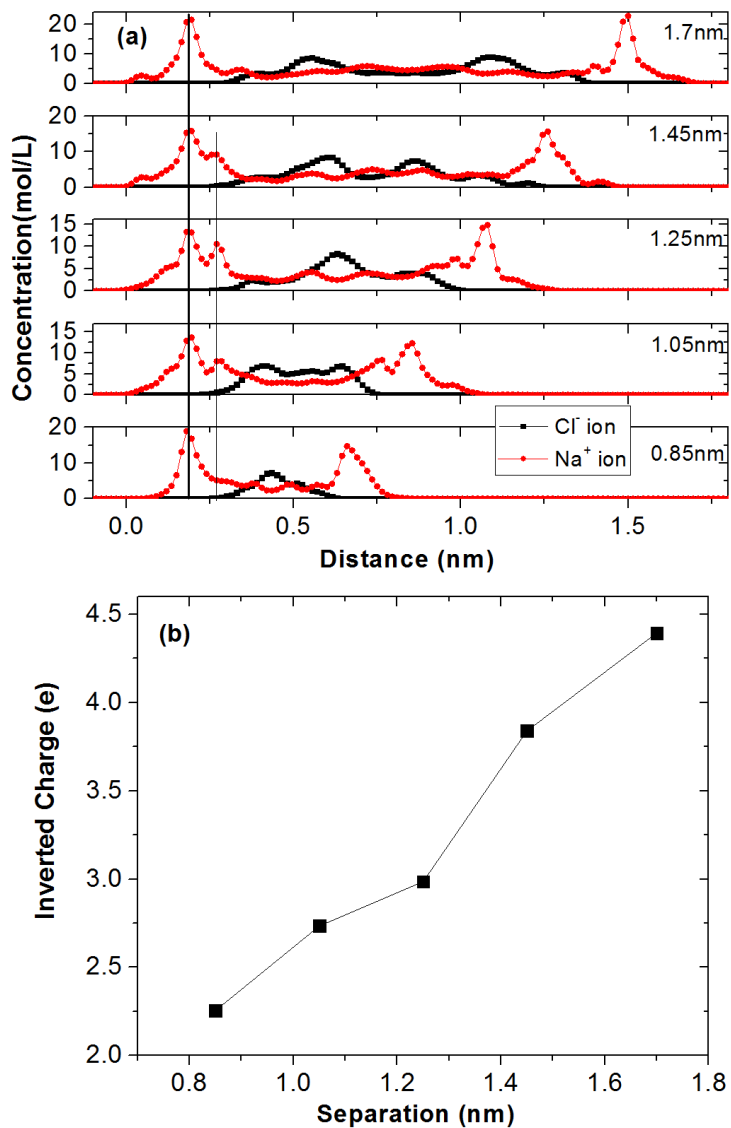

Fig. 8 (a) Ion distributions in the nanoconfined spaces with surface separations from 1.7 to 0.85 nm. The surface charge density in the five cases was the same as −0.15 C/m$^2$. (b) The variation of inverted charge with the surface separation.

From Fig. 9(a), the water molecules show clearly oscillatory distributions. Next to the charged surfaces, the corresponding adsorbed and hydrated layers share the same locations for different cases. As the two surfaces approach each other, the number of water layers in the center region decreases due to the enhanced confinement which is similar to the earlier results in KCl solution confined two mica surfaces.[38] The thickness of each water layer is about 0.25 nm which is close to the diameter of water molecules.[15, 17, 55] By analyzing the coordination number of water molecules nearby the counterions, the hydration performance of $Na^+$ ions was investigated, shown in Fig. 9(b). In the highly confined spaces, $Na^+$ ions have about 5 water molecules[60] in their first hydration shells (0.35 nm in radius, Fig. S11), which is much smaller than the 6.88 found in the bulk case.[5] For different confinement cases, the averaged hydration numbers share almost the same value. We think this is attributed to the similar hydration number distributions near the surface (Fig. S12), i.e. counterions in the main accumulation layers, which account for the majority of ions in the constriction, have nearly the same coordination number. Because the hydration water peak locates farther than the outer Helmholtz layer, as the distance between both surfaces shrinks further, more water molecules will be squeezed out of the confined space than $Na^+$ ions in percent, which causes the hydrated number near $Na^+$ ions much smaller[40]

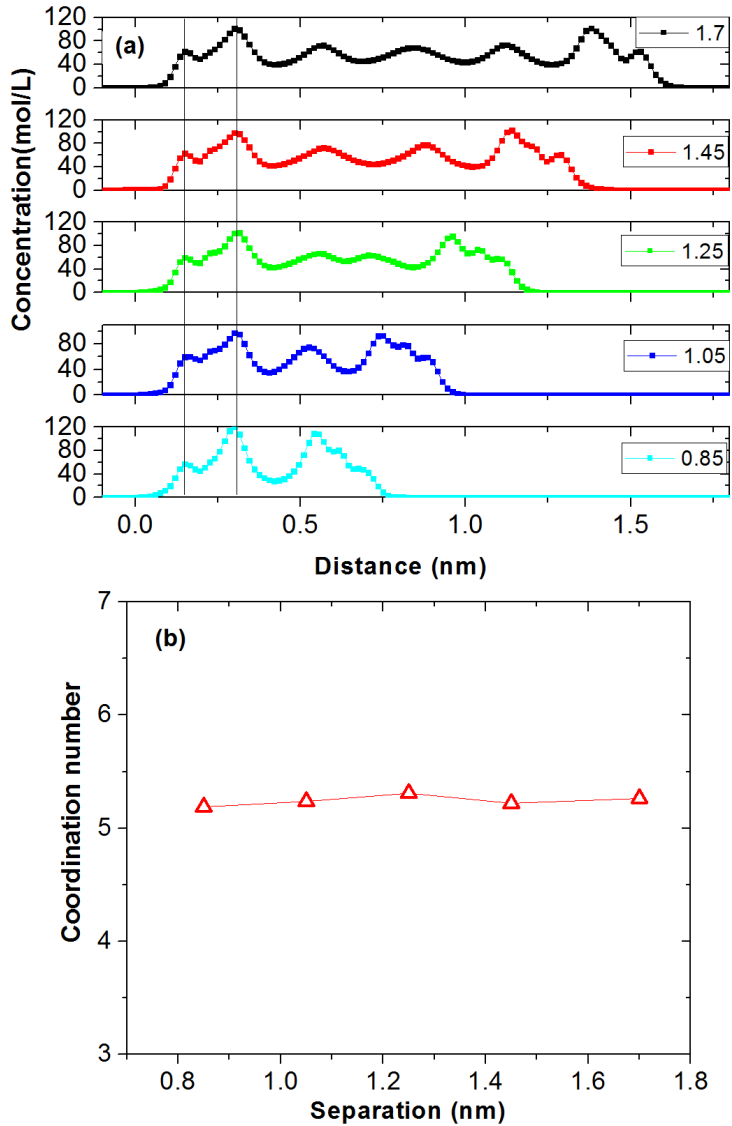

Fig. 9 (a) Water distributions in the nanoconfined spaces with surface separation from 1.7 to 0.85 nm. The surface charge density in the five cases was the same as −0.15C/m$^2$. (b) The variation of coordination number of Na$^+$ ions with the surface separation.

## Conclusion

With the help of the MD simulation model considering charge discreetness and thermal vibration of solid walls, the EDL structure near a much more physically accurate surface was investigated. The obtained ion distributions agree well with the conventional Stern EDL model. When the surface charge density is low, only the outer Helmholtz layer appears near the charged silicon surface. With the surface charge density increasing, the inner Helmholtz layer forms due to the enhanced electrostatic attraction between the surface charges and the counterions. As the surfaces are strongly charged to −0.3 C/m$^2$, a third accumulation layer appears between the inner and outer Helmholtz layers. Due to the strong correlation among the discrete surface charges and counterions, Na$^+$ ions are overcrowded and slight charge inversion appears. With the comparison among the EDL structures in differently concentrated NaCl solutions, the thickness of the EDL is not sensitive to the concentration which is different from the classical theory prediction. The increased counterions in EDL with concentration can cause larger charge inversion and higher peaks of co-ions layers. Because charge inversion happens, Cl$^-$ ions act as counterions to screen the inverted surface charges. The increased co-ions peaks produce larger repulsive EDL force between surfaces which is responsible for the reemergence of the repulsive force found in experiments. In highly concentrated NaCl solutions, the ions in the nanoconfined space show alternating layered distributions. Near strongly charged surfaces, or in highly concentrated solutions or highly confined spaces, this kind of ordered ion distributions becomes much more obvious. In 4.0 M solutions, as the confinement between charged surfaces enhanced, many ions and water molecules are squeezed out. Two adjacent Cl$^-$ ions layers are found to approach each other and merge into one layer accompanying by two adjacent water layers' merging. The coordination numbers of Na$^+$ ions in highly confined spaces are much less than that in the bulk.

## Acknowledgment

The authors thank Zuzanna Siwy, Preston Hinkle, Timothy Plett, William Mann and Crystal Yang at University of California, Irvine for carefully reading and improving the manuscript. The authors acknowledge financial support from the Natural Science Foundation of China (Grant numbers 51435003 and 51445007). This work was partly supported by the Fundamental Research Funds for the Central Universities, the Innovative Project for Graduate Students of Jiangsu Province (Grant No. CXZZ13_0087), the Scientific Research Foundation of Graduate School of Southeast University (YBPY 1504) and the China Scholarship Council (CSC 201406090034). We thank Tianjin computing center for calculations service.

**Supporting Information Available:** This material is available free of charge via the Internet at http://pubs.acs.org.

Tables S1−S3 Numbers of water molecules and ions in the MD systems, Table S4 LJ potential parameters, Figs. S1−S2 Snapshot of the local structures for $Na^+$ ions near charged walls, Fig. S3 MD simulation results with 0.5 M $CaCl_2$ solution, Fig. S4 Coulomb coupling constant at charged surfaces, Fig. S5 Normalized peak values for $Na^+$ and $Cl^-$ ions in different concentration solutions, Fig. S6 Repulsive pressure between two silicon surfaces in NaCl solutions, Fig. S7 Ion concentration distributions under surface charge density $-0.3 C/m^2$, Fig. S8 Debye lengths in monovalent aqueous solutions with different concentrations, Fig. S9 Ion distributions in nanoconfined spaces with different separations, Fig. S10 Integrated charge distributions in nanochannels with the same surface charge density but different separations, Fig. S11 Radial distribution function $g_{Na^+-O}(r)$ in 1.0 M NaCl solution, Fig. S12 Coordination number distributions in nanochannels with the same surface charge density but different separations.

TOC

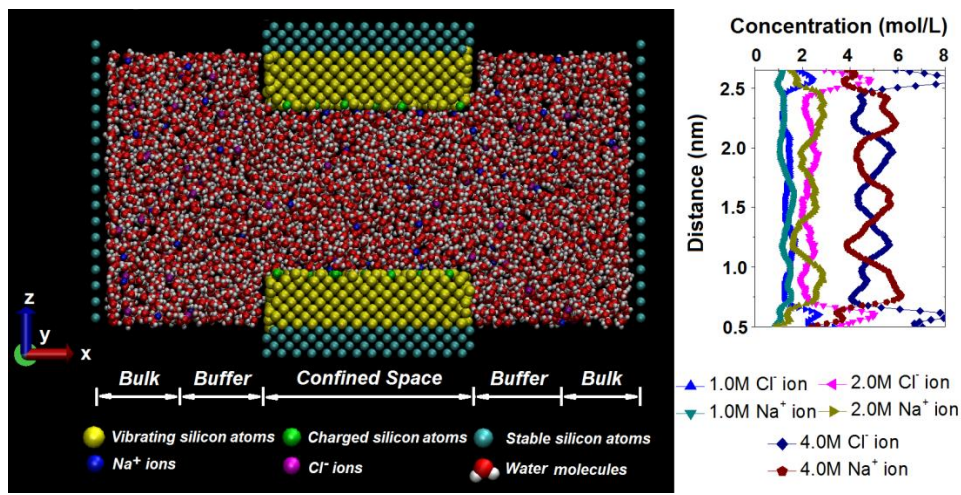